# Heterogeneous adsorption potential of $^3$He in silica aerogel and its influence on magnetic relaxation of $^3$He

**E.M. Alakshin, R.R. Gazizulin, A.V. Klochkov, V.V. Kuzmin, M.S. Tagirov, D.A. Tayurskii**

*Institute of Physics, Kazan (Volga region) Federal University, Kazan, Russia*

*Significant influence of aerogel surface heterogeneity on the processes of $^3$He nuclear magnetic relaxation at temperatures 1.5 – 4.2 K is discovered. This influence appears, for instance, in differences of $^3$He $T_1$ relaxation times for small portion of $^3$He, adsorbed at different temperatures. Binding energy data of $^3$He on the surface of powder silica aerogel obtained experimentally and binding energy lies in the wide range. Adsorbed $^3$He molecules with binding energies 60-250 K play supreme role in processes of nuclear magnetic relaxation of gaseous and liquid $^3$He in aerogel.*

Supreme role of adsorbed $^3$He layer on an aerogel surface in processes of nuclear magnetic relaxation was studied earlier [1,2]. The spin kinetics of $^3$He in the silica aerogel was studied above the Fermi temperature of liquid $^3$He. The magnetic relaxation times $T_1$ and $T_2$ for adsorbed, gaseous, and liquid $^3$He in the 95% porosity aerogel at a temperature of 1.5 K were obtained by means of pulse nuclear magnetic resonance techniques. It was found that $T_1$ in all three cases is proportional to the frequency, whereas $T_2$ is frequency independent. It was shown that the longitudinal relaxation proceeds due to the exchange motion in the solid adsorbed $^3$He film. The intrinsic relaxation mechanisms in the liquid and gas phases are much weaker than the relaxation through the adsorbed surface layer. A theoretical model of relaxation in the adsorbed $^3$He layer, taking into account the filamentary structure of the aerogel, has been proposed.

At present work we report experimental data on heterogeneous adsorption potential of $^3$He in silica aerogel and its influence on magnetic relaxation of $^3$He.

The powder aerogel sample (EMP-SAP (silica aerogel fine powders) EM-POWER CO. LTD (Korea)) was used and sealed leak tight in the glass tube (pyrex) to the gas handling system. The temperature of NMR cell has been controlled by $^4$He vapor pumping and by Allen-Bradley thermometer. The pressure was measured at room temperature part of the gas handling system, using pressure gauge ILMVAC PIZA 111.

The longitudinal magnetization relaxation time $T_1$ of $^3$He was measured by the saturation recovery method using FID signal. The spin-spin relaxation time $T_2$ was measured by Hahn method. The hand made pulse NMR spectrometer has been used (frequency range 3 – 50 MHz). The pulse NMR spectrometer is equipped by resistive electrical magnet with a magnetic field strength up to 1T. Transmission Electron Microscopy images of the sample presented in fig 1. The particles size is about 1- 10 mkm.

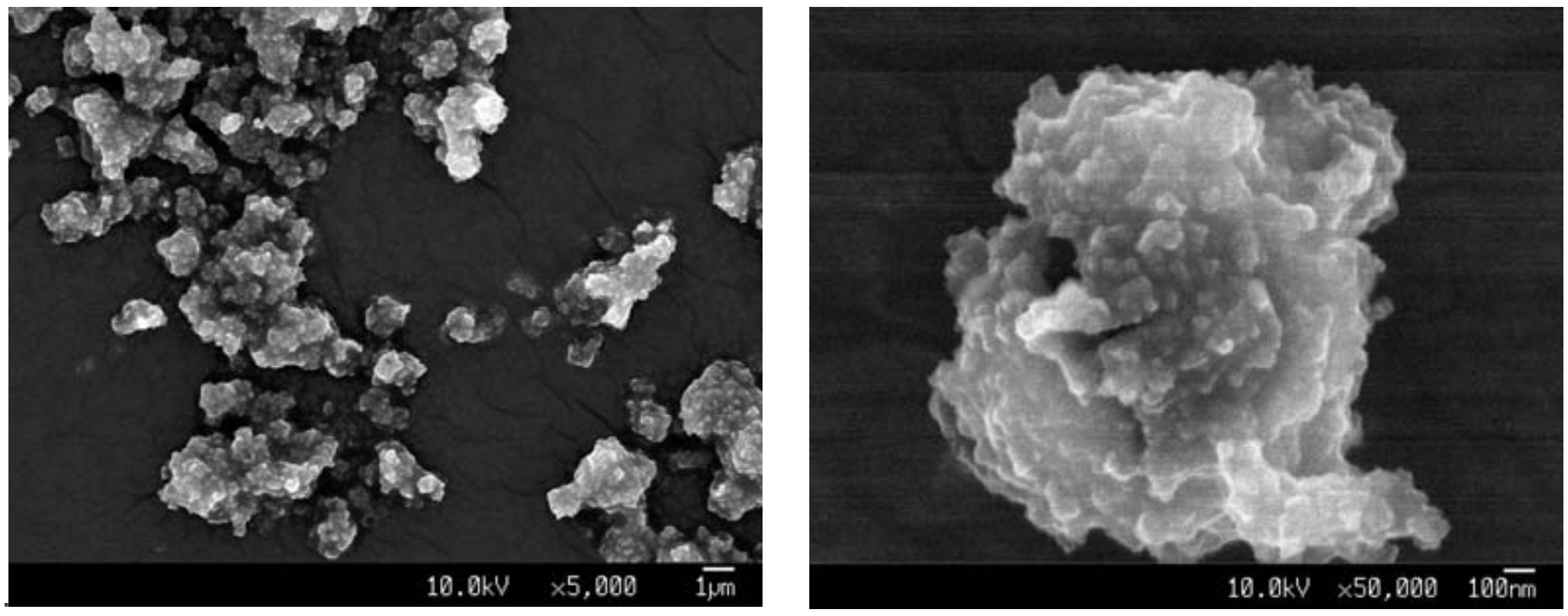

Fig.1. TEM images of powder aerogel sample EMP-SAP.

Adsorption isotherms of $^3$He and $^4$He in an aerogel are presented in fig.2. At the temperature T = 4.2 K small portion of gaseous $^3$He was introduced into the cell, which was further thermalized during 30 min and equilibrium pressure was measured. After each step a new portion of gaseous $^3$He was introduced. The adsorption capacity of complete layer was measured and was equal to 12cm$^3$ STP for $^3$He and 14cm$^3$ STP for $^4$He.

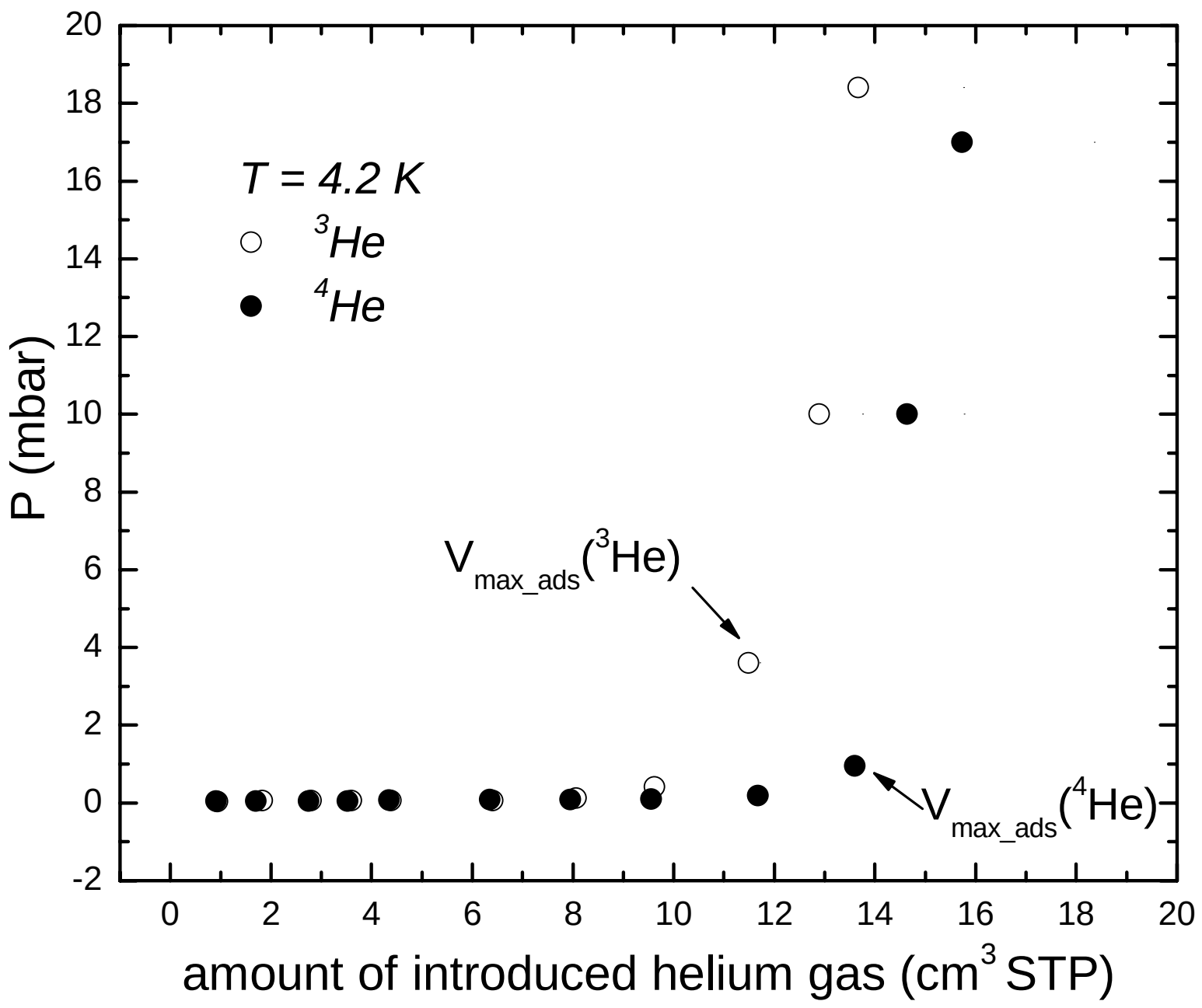

Fig.2. Adsorption isotherms of $^3$He and $^4$He in powder aerogel.

In previous works [1,2] the spin kinetics of $^3$He in the silica aerogel was studied above the Fermi temperature of liquid $^3$He for complete adsorbed layer. In the fig.3. experimental data of $^3$He nuclear longitudinal relaxation time $T_1$ for incomplete adsorbed layer ($V_{ads}$=2cm$^3$ STP) are presented as a function of the temperature.

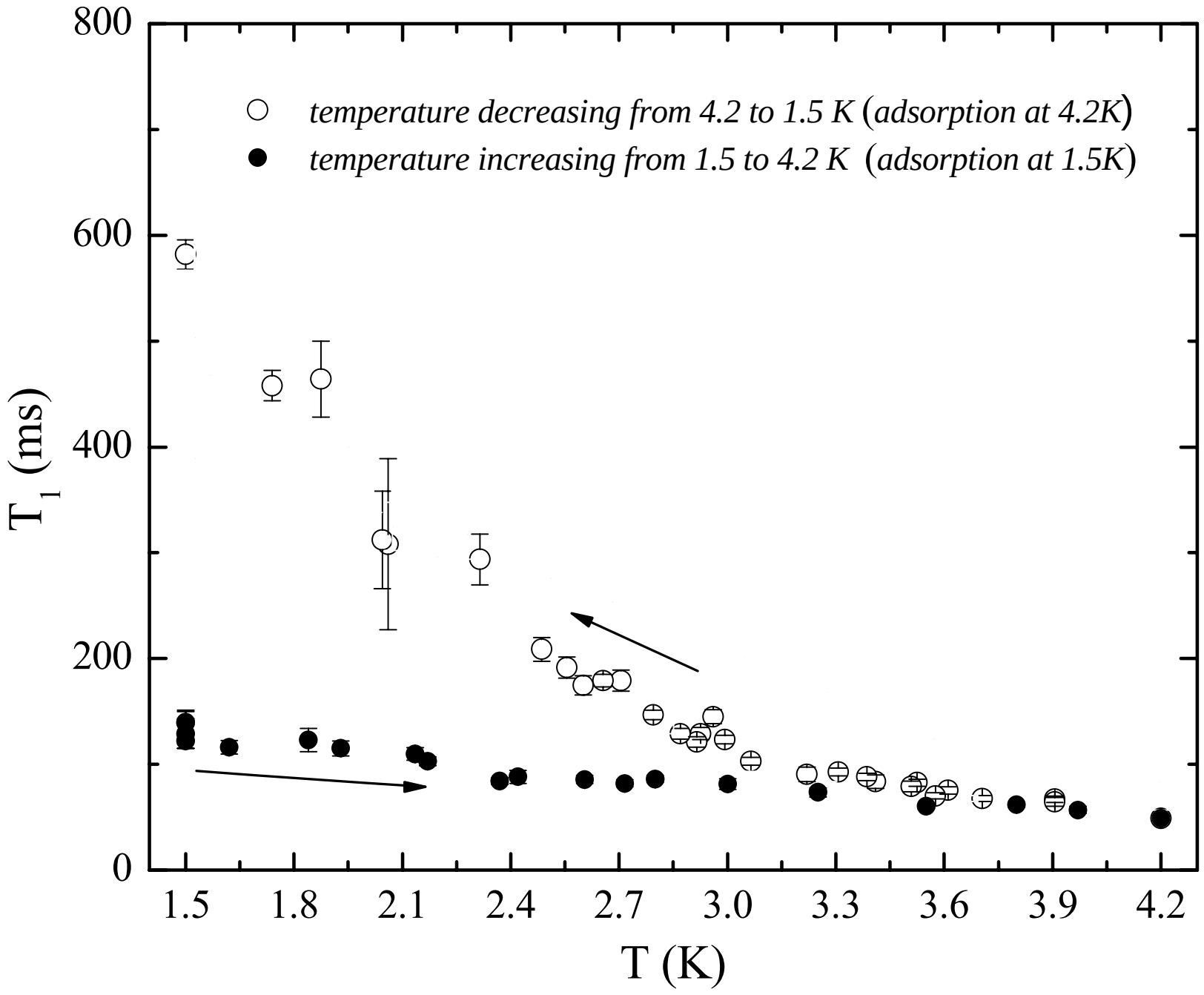

Fig.3. The temperature dependences of $^3$He nuclear longitudinal relaxation time $T_1$ for incomplete adsorbed layer ($V_{ads}$=2 cm$^3$ STP). The adsorbed layer was prepared at two different temperatures (1.5 K and 4.2 K).

The adsorbed layer was prepared at two different temperatures (1.5 K and 4.2 K) and as can be seen from fig.3 the behavior is completely different. It can be explained by taking into account heterogeneity of adsorption potential of $^3$He on an aerogel surface. The adsorption of $^3$He at different temperatures in this case will fill the surface different manner.

The existence of strong temperature dependence of $T_1$ for $^3$He adsorbed on aerogel surface shows that in case of incomplete adsorbed $^3$He layer fast thermal motion and redistribution of $^3$He molecules on the aerogel surface play significant role in nuclear magnetic relaxation of $^3$He.

For obtaining $^3$He adsorption data and calculation binding energies of $^3$He on the aerogel surface following method was used. At room temperature whole gas handling system together with NMR cell was filled by gaseous $^3$He in amount of V($^3$He) = 12 cm$^3$ STP (first experiment) and V($^3$He) = 24 cm$^3$ STP (second experiment). The whole gas handling system consists of “cold part” (capillary in cryostat and NMR cell) and “hot part” (external capillary, calibrating volume and a pressure gauge). The amount of $^3$He was chosen, taking into account fig.2. After equilibrium pressure in whole system was achieved the cryostat begun cool down slowly (1 K / min at temperatures above 10K and 0.05 K / min at low temperatures), using helium flow system.

The cooling process leads to pressure decreasing in the system because of two processes: gas cooling in a “cold part” and adsorption on the aerogel surface. The experimental data are presented in fig.4.

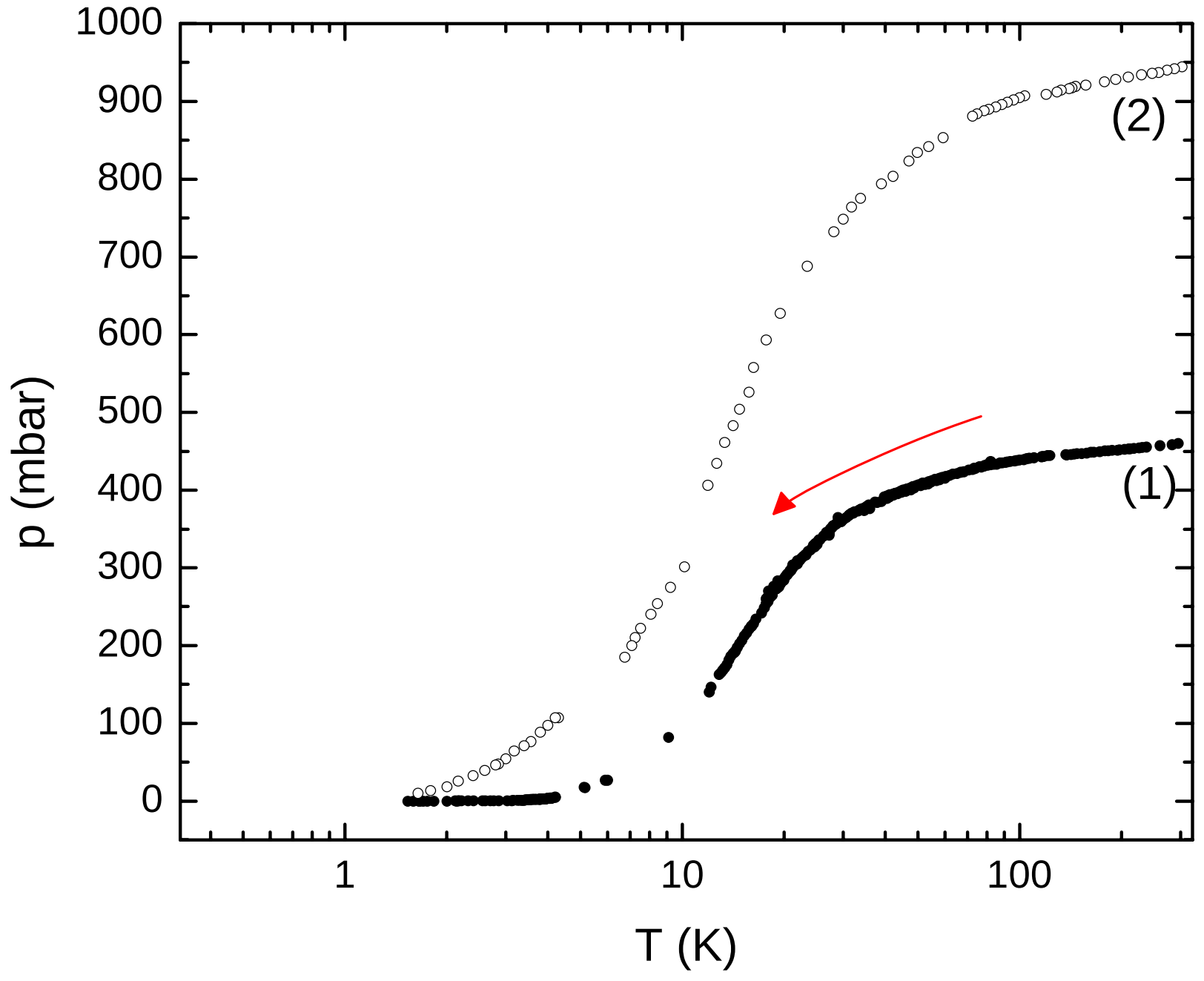

Fig.4. The temperature dependence of $^3$He pressure in the system in two cases: (1) – V ($^3$He) = 12 cm$^3$ STP and (2) – V ($^3$He) = 24cm$^3$ STP.

Taking into account that in the temperature range 70 – 300 K decreasing of the pressure was caused only by cooling down of the $^3$He gas in "cold part" (the adsorption of $^3$He *a priori* is negligible in this temperature range) the dependence in fig.4 in that range was fitted by p=a*T/(1+b*T). This function corresponds to $^3$He gas redistribution between "hot part" and "cold part" during cooling process.

The amount of adsorbed $^3$He was calculated using:

$$M_{ads} = M_0 - M_{hot} - M_{cold}\,, \qquad (1)$$

where $M_0$ – amount of $^3$He in the whole system, $M_{hot}$ – amount of gaseous $^3$He in the "hot part" of the system, $M_{cold}$ – amount of gaseous $^3$He in the "cold part" of the system.

$M_{hot}$ and $M_{cold}$ can be estimated using:

$$M_{hot} = M_0 \cdot p(b/a), \qquad (2)$$

$$M_{cold} = M_0 \cdot p/(b \cdot T), \qquad (3)$$

where a, b – fitting parameters, found earlier, p - pressure, T – temperature.

In fig.5 the results (from eq.1) of described above procedure is presented as a function of $^3$He adsorbed amount versus temperature.

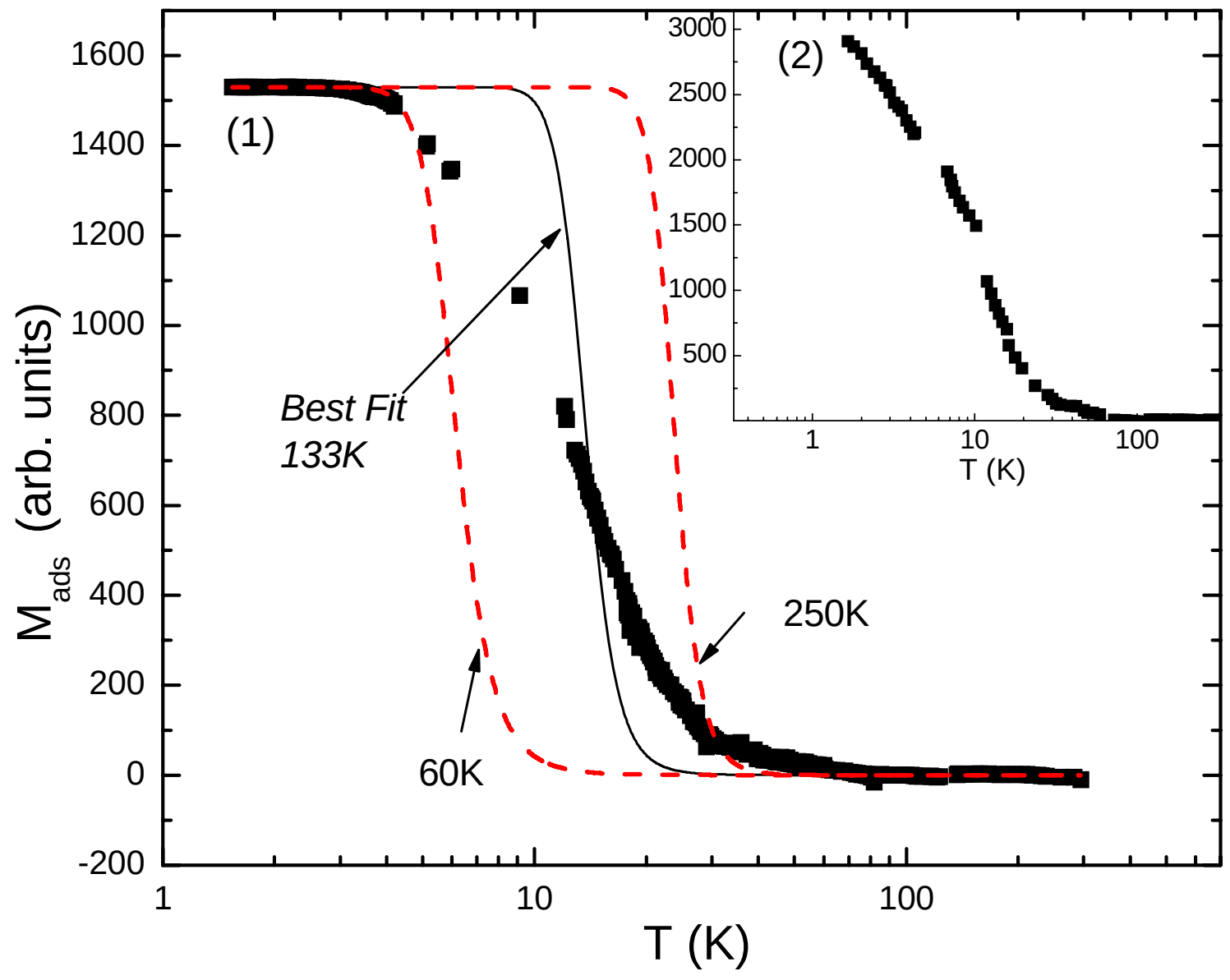

Fig.5. Temperature dependences of $^3$He amount adsorbed in aerogel in two cases: (1) - V($^3$He) = 12 cm$^3$ STP and (2) - V($^3$He) = 24cm$^3$ STP. Dots – experimental data, dash lines – Langmuir model (see text)

The binding energies at V($^3$He) = 12 cm$^3$ STP can be estimated from Langmuir model [3] by eq.:

$$\Theta = p/(p^* + p), \quad (4)$$

where Θ – degree of layer filling

$$p^* = \alpha T^{5/2} / \langle \exp[-H^S/kT] \rangle, \quad (5)$$

where α – constant, T – temperature, $H^S$ – Hamiltonian of $^3$He interaction with an aerogel surface. In the case of low temperatures it can be rewritten as:

$$\langle \exp[-H^S/kT] \rangle \approx \exp[-\varepsilon_m/kT], \quad (6)$$

where $\varepsilon_m$ – $^3$He binding energy on aerogel surface.

The pressure dependences via temperature for calculation of theoretical curves were taken from data in fig.4.

On the fig.5 results of calculations according to (4) are presented for binding energy 60 K, 133 K, 250 K. It is clear, that experimental data could not be described by single binding energy parameter, which proves the existence of heterogeneous adsorption potential. The average binding energy of $^3$He on the surface of powder aerogel EMP-SAP is about 130 K, and the distribution of binding energies lies in the range 60 – 250 K.

On the inset in fig.5 results of $^3He$ adsorption in case of V($^3$He) =24cm$^3$ STP are presented. It can be seen that adsorption continues at lower temperatures, less than 4 K.

The measurements of $^3He$ longitudinal relaxation time $T_1$ during adsorption procedure (fig.4. and fig.5.) in the temperature range 1.5 – 4.2 K are presented in fig.6.

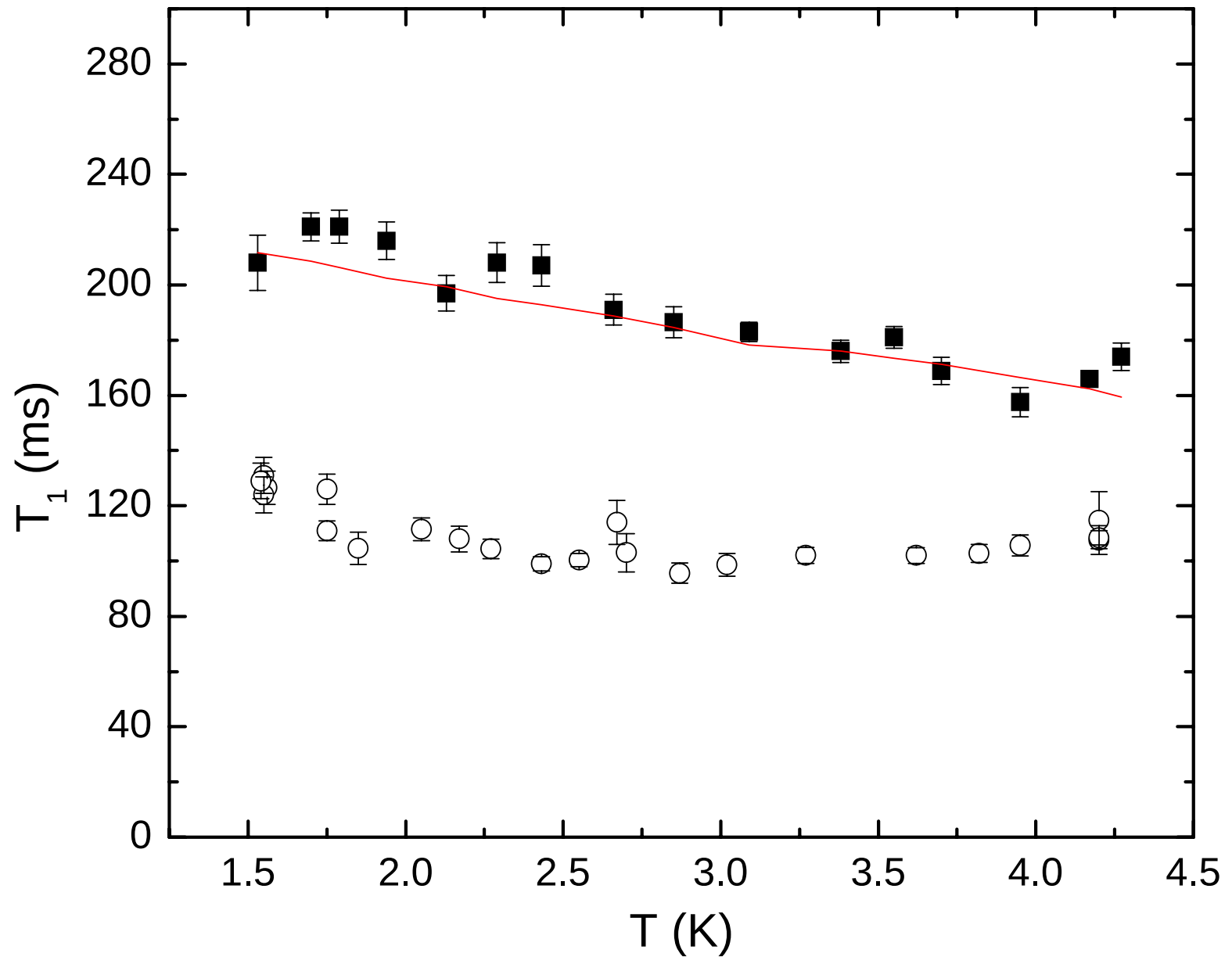

Fig.6. The temperature dependences of $^3He$ longitudinal relaxation time $T_1$ in aerogel: ○ – V($^3$He) = 12 cm$^3$ STP, ■ – V($^3$He) = 24cm$^3$

From fig.6 it is clear, that additional amount of $^3He$ in the cell (case of V($^3$He) = 24cm$^3$) plays role of the load for relaxation process and does not have significant intrinsic relaxation mechanisms. The experimental data can be described, using HR model [4]:

$$T_1=T_{1s}\cdot N_2/N_1 , \qquad (7)$$

where $T_{1s}$ – $^3He$ longitudinal relaxation time in adsorbed (surface) layer, which is almost temperature undependable (fig.6). $N_2$ – amount of $^3He$ molecules in the NMR cell, which depends on temperature. $N_1$ – amount of $^3He$ molecules in adsorbed layer (V($^3$He) = 12 cm$^3$ STP ). The solid line in fig.6 represents calculations, taking into account equation (7) and redistribution of gaseous $^3He$ between “cold part” and “hot part”.

Considering all presented experimental data, significant influence of aerogel surface heterogeneity on the processes of $^3He$ nuclear magnetic relaxation at temperatures 1.5 – 4.2 K was described. For instance, this influence appears in differences of $^3He$ $T_1$ nuclear relaxation times for small portion of $^3He$, adsorbed at different temperatures. Binding energy data of $^3He$ on the surface of powder silica aerogel obtained experimentally and binding energy lays in the range 60 – 250 K. Adsorbed $^3He$ molecules with binding energies in this range play main role in processes of nuclear magnetic relaxation of gaseous and liquid $^3$He in aerogel.

Authors are grateful to Takashi Hattori for TEM measurements of powder aerogel samples.

This work is partially supported by Russian Fund for Basic Research (grant N 09-02-01253).